\documentclass{IEEEcsmag}

\usepackage[colorlinks,urlcolor=blue,linkcolor=blue,citecolor=blue]{hyperref}
\expandafter\def\expandafter\UrlBreaks\expandafter{\UrlBreaks\do\/\do\*\do\-\do\~\do\'\do\"\do\-}
\usepackage{upmath,color}
\usepackage{listings}

\usepackage{tikz}

\jvol{XX}
\jnum{XX}
\paper{8}
\jmonth{September/October}
\jname{IEEE Micro (Preprint)}
\jtitle{IEEE Micro (Preprint)}
\pubyear{2024}

\begin{document}

\sptitle{Contemporary Industry Products}

\title{LPU: A Latency-Optimized and Highly Scalable Processor for Large Language Model Inference}

\author{Seungjae Moon\textsuperscript{1}, Jung-Hoon Kim\textsuperscript{2}, Junsoo Kim\textsuperscript{1}, Seongmin Hong\textsuperscript{1}, Junseo Cha\textsuperscript{1}, Minsu Kim\textsuperscript{2}, Sukbin Lim\textsuperscript{1}, Gyubin Choi\textsuperscript{1}, Dongjin Seo\textsuperscript{1}, Jongho Kim\textsuperscript{1}, Hunjong Lee\textsuperscript{1}, Hyunjun Park\textsuperscript{1},  Ryeowook Ko\textsuperscript{1}, Soongyu Choi\textsuperscript{1}, Jongse Park\textsuperscript{2}, Jinwon Lee\textsuperscript{1}, Joo-Young Kim\textsuperscript{1}}
% \author{Second Author Jr.}
% \affil{Company, City, (State), Postal Code, Country}
\affil{\textsuperscript{1}HyperAccel, Seoul, 06247, South Korea, \textsuperscript{2}Korea Advanced Institute of Science Technology, Daejeon, 34141, South Korea}

\markboth{Contemporary Industry Products}{Contemporary Industry Products}

\begin{abstract}\looseness-1The explosive arrival of OpenAI's ChatGPT has fueled the globalization of  large language model (LLM), which consists of billions of pretrained parameters that embodies the aspects of syntax and semantics. HyperAccel introduces latency processing unit (LPU), a latency-optimized and highly scalable processor architecture for the acceleration of LLM inference. LPU perfectly balances the memory bandwidth and compute logic with streamlined dataflow to maximize performance and efficiency. LPU is equipped with expandable synchronization link (ESL) that hides data synchronization latency between multiple LPUs.  HyperDex complements LPU as an intuitive software framework to run LLM applications. LPU achieves 1.25 ms/token and 20.9 ms/token for 1.3B and 66B model, respectively, which is 2.09$\times$ and 1.37$\times$ faster than the GPU. LPU, synthesized using Samsung 4nm process, has total area of 0.824 mm\textsuperscript{2} and power consumption of 284.31 mW. LPU-based servers achieve 1.33$\times$ and 1.32$\times$ energy efficiency over NVIDIA H100 and L4 servers, respectively.
\end{abstract}

\maketitle

\chapteri{T}he fundamental goal of AI is to create human-like intelligence. Conventional AI has reached a level of human ability to enable data analysis, decision-making, and personalization. It is now advancing at a remarkable pace, even in domains once thought to be uniquely human, such as creativity, but it was yet to replicate the creativity of humans. Generative AI, or GenAI, has made a recent breakthrough with the transformer models (e.g., GPT and LLaMA) that are capable of creating original textual and image contents with high sophistication.
Specifically, GenAI software platforms based on the large language models (LLM) with multi-billion parameters, such as OpenAI ChatGPT and Google Bard, are in the forefront of revolutionizing the usage of AI.
The growing efforts to commercialize the LLM models and effectively support these advanced AI platforms highlight the critical need for the development of specialized inference hardware in datacenters.

Introduced by Vaswani et al., LLM model inference is based on the transformer decoder, in which the inputs have limited batching capabilities and require sequential processing\textsuperscript{1}. Since relatively small inputs need to be inferred with large model parameters, the inference incurs memory bottleneck and requires efficient processing of the system's memory bandwidth. Moreover, scalability becomes significant as the ever-increasing compute and memory requirements of LLMs demand multiple devices and communication between them. At the application level, each user makes individual requests and expects the generated output with minimal wait time, making it crucial to have a hardware platform that reduces inference latency. The predominant hardware for inference, GPU, underperforms for GenAI workloads because it undergoes low hardware utilization for small-batch inputs and high communication overhead during synchronization. Therefore, a new class of processor is required that targets the memory-intensive GenAI workloads.

In addition to performance, efficiency and usability are important factors in evaluating an inference hardware. Maintaining efficiency across every GenAI application is difficult because each application requires different sizes of LLM. Although larger LLM may generate superior response for a given user context, the power and cost overhead of running such model may not be suitable. For instance, the service-level agreement (SLA) is less strict for LLM applications used for light interaction with the user (e.g., chatbot), in which smaller LLMs may be adequate, whereas an application that conducts a search or analysis from a broad context may require larger LLMs as discussed by Chowdhery et al\textsuperscript{2}. However, hardware that considers the acceleration of larger models likely undergoes inefficiencies for smaller models. Therefore, an architecture that fully leverages the memory bandwidth to yield maximum performance regardless of the model size would be the most efficient. The hardware must also be accompanied by a comprehensive software framework for fast speed-to-market of these various LLMs. A compiler that is both model-and-hardware-aware and automated to output the prerequisite data, such as memory mapping and instructions, is necessary. The runtime software also must adhere to the available API, such as HuggingFace, so that a hardware solution can be easily integrated to run LLM applications.

In this paper, we propose HyperAccel's \textbf{latency processing unit (LPU)}, a latency-optimized and highly scalable architecture that accelerates large language model inference for GenAI. The key contributions of the LPU are as follows:

\begin{itemize}
\item LPU introduces streamlined hardware that maximizes the effective memory bandwidth usage during end-to-end inference regardless of the model size to achieve up to 90\% bandwidth utilization for high-speed text generation. It also consists of \textbf{expandable synchronization link (ESL)} that hides bulk of the data synchronization latency in a multi-device system to achieve near-perfect scalability, or 1.75$\times$ speedup for doubling the number of devices.
\item We propose \textbf{HyperDex}, a software framework that enables automated compilation of prerequisite data based on LLM specifications. It also provides a runtime environment based on widely used HuggingFace API for seamless execution of GenAI applications on LPU hardware.
\item LPU achieves 1.25 ms/token for OPT 1.3B, and two LPUs achieve 20.9 ms/token for OPT 66B, which is 2.09$\times$ and 1.37$\times$ faster than GPUs with equal device count. The LPU-based ASIC implemented using 4nm process consumes only 0.824 mm\textsuperscript{2} in area and 284.31 mW in power.
\item We showcase HyperAccel \textbf{Orion}, an LPU-equipped server system that is ready to run GenAI application in cloud and edge datacenters. Orion achieves 1.33$\times$ and 1.32$\times$ energy efficiency over the state-of-the-art NVIDIA H100 and L4 GPU server solutions, respectively.
\end{itemize}

\section{Background and Motivation}

%%%%%%%%%%%%%%%%%%%%%%%%%%%%%%%%%%%%%%%%%%%%%%%%%%
% Figures
%%%%%%%%%%%%%%%%%%%%%%%%%%%%%%%%%%%%%%%%%%%%%%%%%%
\begin{figure}[t] 
\vspace{0in}
\centering
\footnotesize
\includegraphics[width=2.8in]{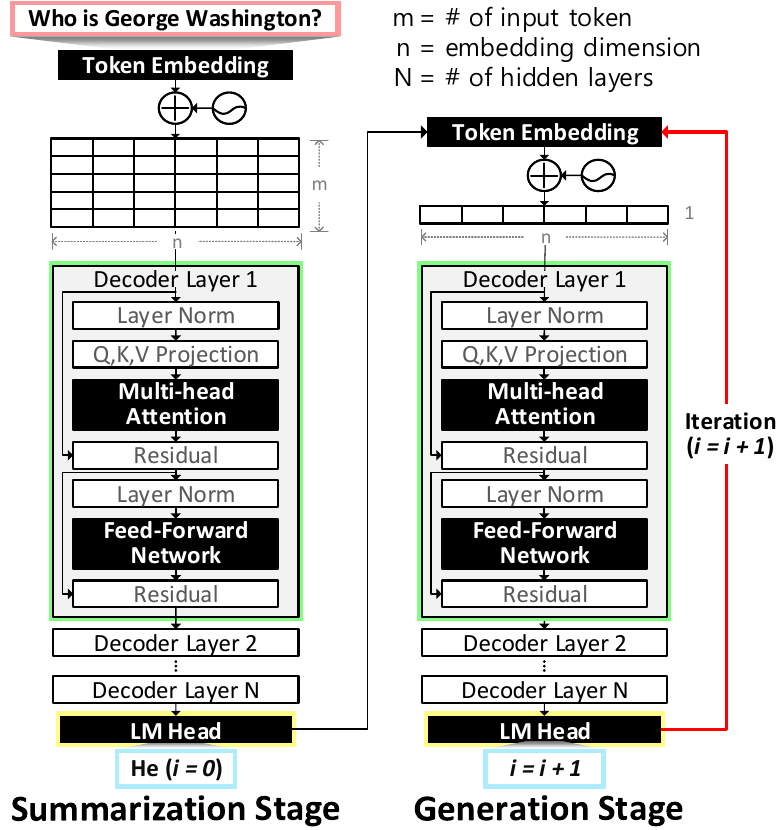}
\vspace{-0.05in}
\caption{Structure of large language model.}
\label{fig-transformer} 
\vspace{-0.20in}
\end{figure}
%%%%%%%%%%%%%%%%%%%%%%%%%%%%%%%%%%%%%%%%%%%%%%%%%%

% $m$ refers to the number of input token length, and $n$ is the embedding dimension of the embedding vectors..
% m = # of input token
% n = embedding dimension
% N = # of hidden layers

\subsection{Structure of Large Language Model}
Large language model inference is largely divided into summarization and generation stage as shown by Wang, Zhang, and Han\textsuperscript{3}. Figure \ref{fig-transformer} shows the overview of the LLM inference.
LLM inference begins with the summarization stage, in which the input to the decoder layer is a matrix that represents the token embedding of the user context (e.g., statement or question). The matrix is inputted to a series of decoder layers, which is based on the transformer decoder. The final decoder layer outputs the captured features from the input context. The result of the final decoder layer enters the language modelling head, or LM head. LM head converts the features into logits that score the candidate tokens from the dictionary based on their likelihood of being appropriate in the given context. With a single execution of the summarization stage, the first output token ($i=0$) is produced. The first output token then enters the generation stage. Only Keys and Values from the masked multi-head attention are transferred to the next stage as activations that hold contextual information about the previous token.
In the generation stage, the input to the decoder layer is guaranteed to be a single embedding vector. The processes are repeated to autoregressively output the next output token ($i=i+1)$. The generation stage iterates until the end of sequence token is reached.

Out of all the processes, the most computationally dominant process is the decoder layer. Furthermore, masked multi-head attention and feed-forward network operations within the decoder layer account for 90.7\% of the total inference time for LLM with 7 billion parameters (e.g., Llama 7B). Both operations require the input to be multiplied by a weight matrix, and the input is a matrix in the summarization stage and a vector in the generation stage. Since more iterations are required in the generation stage, an architecture that is optimized for vector-matrix multiplication is necessary for achieving maximum performance. 

Sequential characteristic of LLM inference requires a constant access to new parameters with minimal reuse, which indicates that LLM inference undergoes significant memory bottleneck. To resolve the memory bottleneck, sufficiently fast memory is required (i.e., high bandwidth memory or HBM), but more importantly, the efficient use of the given bandwidth is directly proportional to performance.

%%%%%%%%%%%%%%%%%%%%%%%%%%%%%%%%%%%%%%%%%
\subsection{Diversity of Large Language Model}
The size of large language model is diverging. The demand for increasingly flexible and accurate LLMs has initiated a competitive push towards larger LLMs, some boasting up to a trillion parameters, whereas the demand for more manageable models have reduced the model size to single-digit billion parameters with optimization efforts in prompt engineering, domain-specific training, and various quantization methods. Depending on the environment, device with different hardware specifications, especially memory bandwidth, may be adopted due to different SLA and price considerations. For instance, a device with lower memory bandwidth may satisfy if SLA is less strict or budget is limited. 

In order to meet the requirements of both diverse-sized large language models and scalable hardware, a software framework that effectively bridges them is essential.
Wolf et al. identifies that models are typically defined and deployed to the acceleration hardware using software frameworks such as HuggingFace Transformers, PyTorch, and TensorFlow in the LLM ecosystem\textsuperscript{4}.
The software framework performs optimization by considering several system-wide configurations such as model size, budget, SLA, and hardware specifications, to meet the requirements.
These frameworks hide complex details of the underlying acceleration hardware, enabling AI model architects and application developers to leverage software framework to describe their models and integrate applications into the hardware, respectively.
%

%%%%%%%%%%%%%%%%%%%%%%%%%%%%%%%%%%%%%%%%%

% FIGURE: GPU weakness graphs (low utilization, high power, low scalability)
%%%%%%%%%%%%%%%%%%%%%%%%%%%%%%%%%%%%%%%%%%%%%%%%%%
% Figures
%%%%%%%%%%%%%%%%%%%%%%%%%%%%%%%%%%%%%%%%%%%%%%%%%%
\begin{figure}[t] 
\vspace{0in}
\centering
\footnotesize
\includegraphics[width=2.8in]{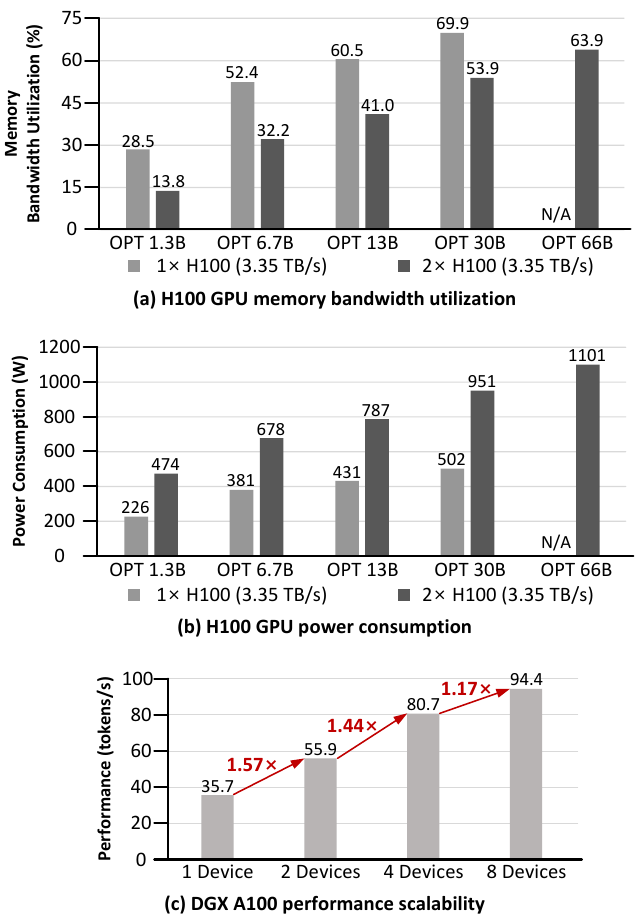}
\vspace{-0.05in}
\caption{GPU analysis when running LLM inference.}
\label{fig-motive} 
\vspace{-0.20in}
\end{figure}
%%%%%%%%%%%%%%%%%%%%%%%%%%%%%%%%%%%%%%%%%%%%%%%%%%

\subsection{Inefficiency in Conventional Hardware}
\label{gpu_inefficiency}
\textbf{Bandwidth.} Although modern GPU features substantial memory bandwidth and computation power, utilizing the full bandwidth of a GPU for LLM inference can be challenging due to the way GPU is designed. For instance, the architecture of a NVIDIA GPUs are optimized for parallel processing, where many threads execute simultaneously across multiple cores. Especially in the generation stage, GPU cannot effectively route the incoming bandwidth to a single core that requires computation with a single vector at a time, which causes underutilization of both compute cores and memory bandwidth as mentioned by Hong et al.\textsuperscript{5} This innate problem is more pronounced in smaller models due to even smaller operands (i.e., input and activations). According to Neubig et al. and Kitaev et al., a myriad of software techniques, such as in-flight batching, Key-Value caching, and other algorithmic optimizations have been proposed to raise the efficiency of GPUs\textsuperscript{6, 7}. Despite these efforts, the utilization when processing real inference workloads is bounded by the physical limitations, which leads to inefficiencies when deploying GPUs in practice. On average, NVIDIA H100 GPU achieves as low as 28.5\% bandwidth utilization for the smaller OPT 1.3B model but up to 69.9\% for the larger OPT 30B model. Figure \ref{fig-motive}(a) shows the bandwidth utilization of running LLM of various sizes.

\textbf{Power.} The high memory bandwidth and compute power of GPU comes with high power consumption. Since LLM inference is a memory-intensive workload, the power consumed during data transfer from memory translates to performance. However, the high operating frequency and the number of unused compute cores result in unnecessary power consumption along with other power-hungry peripherals. For running OPT 66B model, two NVIDIA H100 GPUs consume an average of 1101 W. Figure \ref{fig-motive}(b) shows the power consumption of running LLM of various sizes. 

\textbf{Scalability.} The growing size of LLMs have led to the requirement of multiple devices. Since the data precision of standard LLM models are half-precision floating-point (FP16), the memory requirement of LLM is approximately number of model parameter multiplied by two bytes. For instance, 66B model requires 132 GB and additional 5 GB for storing Key-Value. Since 137 GB exceeds the HBM capacity of NVIDIA H100 GPU with 80 GB, two H100s are required. Therefore, an effective communication between devices is essential for efficient LLM inference of large models. NVIDIA GPUs support NVLink, a flagship direct GPU-to-GPU interconnect that transfers data up to 900 GB/s. Despite this high-speed interconnect, the data synchronization overhead in tensor parallelism is significant because the computation is stalled during the communication. Figure \ref{fig-motive}(c) shows the scalability of NVIDIA DGX A100 with third generation NVLink (600 GB/s) running GPT3 20B inference. DGX A100 achieves only an average of 1.38$\times$ speedup when doubling the number of devices. The scalability result is based on the released benchmark result of the NVIDIA FasterTransformer (FT) library. 
For the most efficient processing, domain-specific architecture for LLM inference with high bandwidth and core utilization, low power overhead, and high scalability is required.

\section{Latency Processing Unit}

\begin{figure}[t] 
\vspace{0in}
\centering
\footnotesize
\includegraphics[width=3.0in]{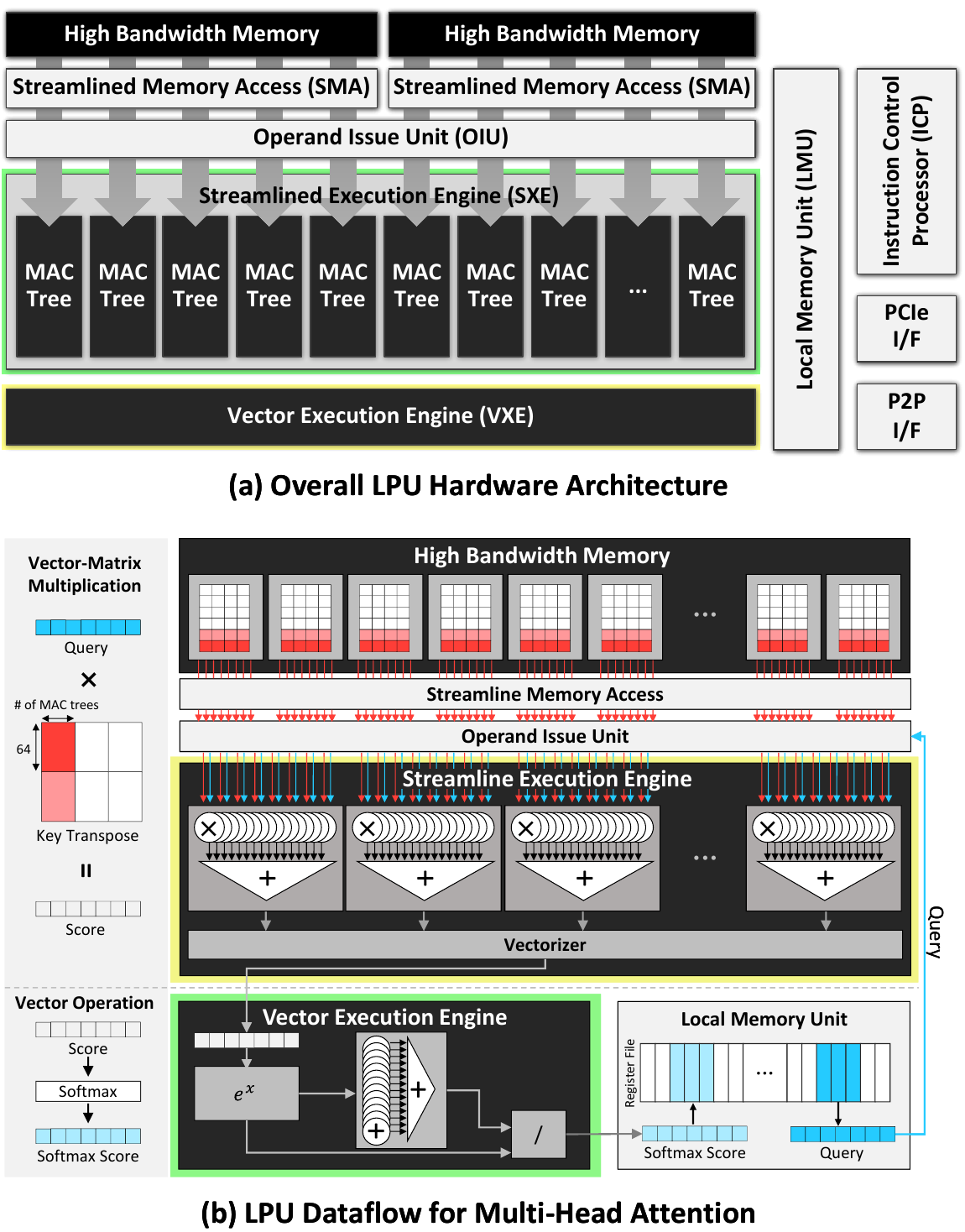} 
\vspace{-0.2in}
\caption{LPU hardware architecture and its dataflow.}
\label{fig-architecture} 
\vspace{-0.2in}
\end{figure}

The \textbf{latency processing unit (LPU)} architecture consists of streamlined hardware, custom instruction set, and LLM-specific dataflow for high-speed LLM inference. The LPU architecture is shown in Figure \ref{fig-architecture}(a). 

\subsection{Hardware Architecture}

%%%%%%%%%%%%%%%%%%%%%%%%%%%%%%%%%%%%%%%%%%%%%%%%%%
% Figures
%%%%%%%%%%%%%%%%%%%%%%%%%%%%%%%%%%%%%%%%%%%%%%%%%%
%%%%%%%%%%%%%%%%%%%%%%%%%%%%%%%%%%%%%%%%%%%%%%%%%%

\textbf{Streamlined Memory Access (SMA)} is a specialized DMA that connects all HBM channels to the execution engines to transfer FP16 data at maximum bandwidth. It preloaded with memory instructions that sends continuous read requests for weights and occasional write requests for Key-Value write. Our hardware-aware memory mapping removes the need for any data reshaping or switching in the SMA. Matrix transpose is required in the attention mechanism, but SMA utilizes the strobe signal with an algorithmic approach that writes to specific memory location so that the data is naturally transposed when read without adding latency overhead. Since a number of compute units are placed to exactly match the total HBM bandwidth, SMA streams the data received at maximum burst size to the execution engines with minimum stalling. The streamed data are parameters for vector-matrix execution (e.g., weight, bias) and other vector-related operations (e.g., gamma/beta, embedding). 

\textbf{Operand Issue Unit (OIU)} arbitrates the streamed data (i.e., first operand) from the SMA and the input (i.e., second operand) from the on-chip register file before issuing them to the execution engines. Based on the compute instructions, the OIU generates microcodes that configure the execution engines and decide the destination engine for the operands. Appropriate operands are prefetched, ready to be issued to the execution engines immediately.

\textbf{Streamlined Execution Engine (SXE)} is LPU's main computing hardware with custom low-latency multiply-accumulate (MAC) trees, which are primarily designed to execute vector-matrix multiplication for multi-head attention and feed-forward network operations. Each MAC tree consists of the following features: 1) The preprocessing of the operands based on the exponent and mantissa of the larger floating-point operand enables the fixed-point multiplication and accumulation to reduce the logic area. 2) The fixed-point adder tree for mantissa utilizes a Wallace tree for high-speed addition via parallelization. The SXE has $l$ number of MAC trees, each operating on a set of $v$ vector elements. To fully utilize the memory bandwidth, the incoming bandwidth and the bandwidth of the MAC tree (i.e., $l\times v\times 2\,B\times freq$) must be equal. Note that $freq$ refers to the operating frequency of the hardware. Assuming the target frequency is predetermined, number of MAC trees is chosen to match the memory speed. Moreover, SXE is superpipelined to constantly receive operands, which increases the throughput and thus lowers the overall latency. It also supports model-specific operations such as rotary positional embedding and nonlinear activation functions.

\textbf{Vector Execution Engine (VXE)} executes vector operations, such as token embedding, softmax, normalization, and residual, with custom low-latency ALU. VXE also contains a sampler that sorts logits and selects an output token based on temperature, top-p, and top-k values. Since these vector operations occur less frequently, we reduce the fan-in from the OIU to this path to decrease the hardware resource with negligible performance loss.

\textbf{Instruction Control Processor (ICP)} is a RISC processor that controls the overall execution flow of the LPU. The ICP primarily fetches LPU instructions from the instruction buffer and dispatches them. It also executes basic RISC-type instructions, such as branch and jump, based on the control registers (e.g., token and layer number) for iterative and conditional logic. The dispatcher in the ICP is entirely independent of other LPU modules, so the instructions are continuously prefetched in the other modules for minimum control interference. Moreover, the internal scheduler supports the out-of-order execution of SXE and VXE for improved latency and hardware utilization, and a scoreboard is designed to handle data hazards. 

\textbf{Local Memory Unit (LMU)} is a multi-bank register file with scalar-vector segregation for fast high-bandwidth access to one of the operands. It is also multi-port to support simultaneous read to the OIU and write from the writeback of the execution engines.

\subsection{LPU Dataflow}
 The LPU adopts the output stationary dataflow, in which the activation vector is reused, and weights are streamed to execute vector-matrix multiplication. Rectangular tiles with the length equal to the vector dimension and the width equal to the number of MAC trees are loaded from memory each cycle. The tiles are accessed in the vertical direction, which reduces the number of partial sum buffers and simplifies the control compared to other directions (e.g., horizontal and zigzag) because a set of dot products is guaranteed to be finished before the next set begins.The tile is memory mapped so that the data requested from the SMA can be directly wired as inputs to the MAC trees to fully utilize the SXE during the time vector-matrix multiplication is being executed. The parameters are also mapped to stream to VXE for vector-vector operations. Figure \ref{fig-architecture}(b) shows an example of the LPU dataflow when executing multi-head attention. The memory mapped Key stored in the HBM is read to the SMA based on the tiling. The Key and the corresponding portion of the Query stored in the LMU are multiplied in SXE using the custom MAC trees to generate a Score. The Score is then sent to VXE to execute the softmax operation while the remaining tile of Key enters the SXE. During the parallel execution of SXE and VXE, the SMA continuously reads the remaining Key from the HBM, and the ICP concurrently calculates the new addresses for the next memory read/write.

%%%%%%%%%%%%%%%%%%%%%%%%%%%%%%%%%%%%%%%%%%%%%%%%%%
% Table
%%%%%%%%%%%%%%%%%%%%%%%%%%%%%%%%%%%%%%%%%%%%%%%%%%
\begin{table}[t]
\caption{LPU instruction set architecture.}
\centering
\includegraphics[width=3.0in]{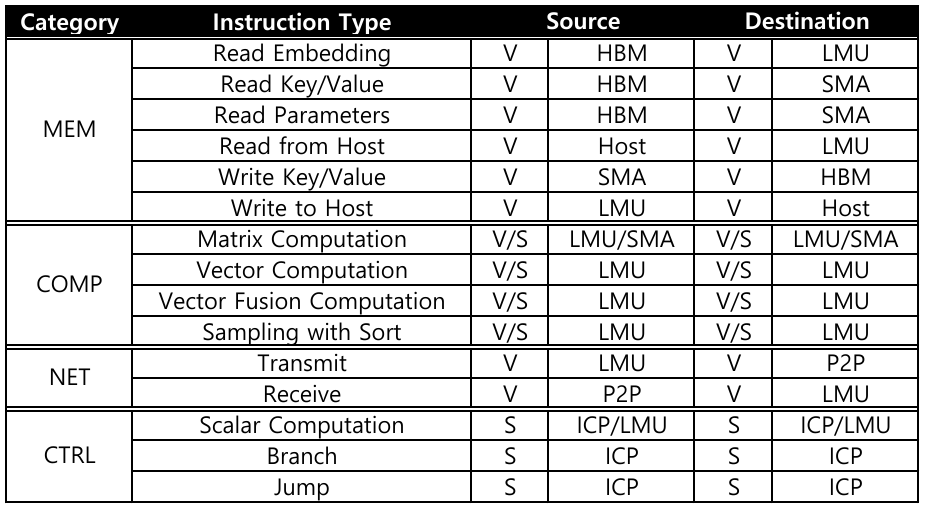}
\label{tab-isa}
\vspace{-0.20in}
\end{table}
%%%%%%%%%%%%%%%%%%%%%%%%%%%%%%%%%%%%%%%%%%%%%%%%%%

\subsection {Instruction Set Architecture}
\label{ISA}
The LPU operates based on a custom instruction set architecture (ISA). Table \ref{tab-isa} shows a shortlist of the ISA of LPU. The ISA is largely divided into direct memory access (MEM), compute (COMP), network (NET), and basic RISC-type instructions (CTRL). MEM accommodate memory read/write of different parameters (e.g., weight, Key, Value, embedding and normalization data) for data transfer. COMP handles simple vector arithmetic along with more complex fusion operations for computation. NET handles transmit/receive of partial results for synchronization. CTRL moves and calculates the program counter and addresses based on scalar register data for controlling the processor. The processes of reading the weight, executing the calculation, and writing the activation are executed concurrently for independent instruction sets and in a streamlined manner for dependent instruction sets, to fully utilize the bandwidth, which translates to enhanced performance.

\section{Expandable Synchronization Link}

With the increase in model size, running LLM on a single device is becoming challenging. The intra-layer model parallelism, which distributes both the model parameters and computational load across multiple devices to execute partial vector-matrix multiplications, has become the standard approach. Since the entire resulting vector is required before a subsequent multiplication begins, the compute units are stalled until the synchronization of data is completed, which causes a significant overhead. A larger model requires even more data and devices to communicate with per synchronization, which would further increase the overhead. To address this issue, we develop \textbf{expandable synchronization link (ESL)}, a peer-to-peer (P2P) communication technology that performs data synchronization with latency hiding. ESL also supports reconfiguration to enable optimal network configuration for the given workload.

\subsection{Expandable Network} 

\begin{figure}[t] 
\vspace{0in}
\centering
\footnotesize
\includegraphics[width=2.8in]{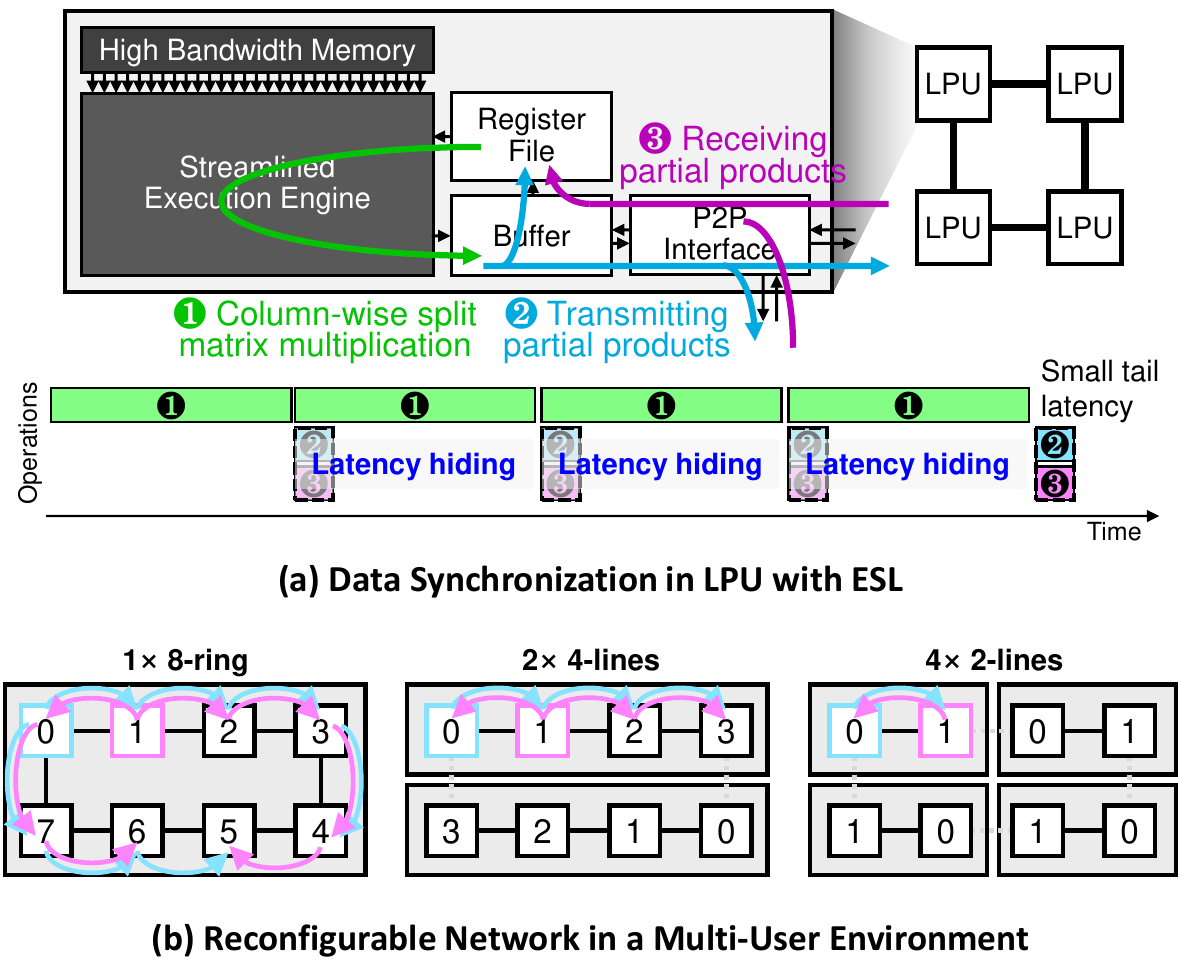} 
\vspace{-0.10in}
\caption{Dataflow and timeline of data synchronization in LPU with expandable synchronization link.}
\label{fig-esl} 
\vspace{-0.2in}
\end{figure}

ESL primarily focuses on fostering high scalability that ensures linear performance improvement with the expansion of the number of LPU devices. We devise a custom ESL protocol that effectively overlaps vector-matrix multiplication with synchronization, thereby hiding the communication latency. In a typical processor that adopts model parallelism (e.g., tensor parallelism), the communication follows after the computation, and vice versa, leading to the inclusion of the entire communication latency in the overall latency. In ESL, vector-matrix multiplication is first divided into smaller column-based tasks, in which the result matches the bitwidth of the P2P interface. The destination of the partial products from SXE is assigned to be a temporary buffer instead of the register file. From the buffer, the supporting ESL dataflow enables the immediate transmission of the partial products to the peer devices while the next operation is ongoing. The dataflow also includes the runtime arbitration between the partial products received from the peer devices and written back from its own SXE. Since computing, transmitting, and receiving can all be done concurrently, new computation and the communication pertaining to the previous computation always overlap. This overlap hides all of the communication latency except for a small tail latency. Figure \ref{fig-esl}(a) shows the ESL dataflow and the operation timeline. For cases in which two vector-matrix multiplications happen sequentially (e.g., fully connected layer (FC) 1 followed by FC layer 2 in the feed-forward network), even the tail latency of the synchronization is hidden. 
% for the first vector-matrix multiplication because the second partial vector-matrix multiplication can begin without synchronizing the last partial results of the first vector-matrix multiplication. Because this pattern occurs in half of the instances where synchronization is required, the ESL has near-zero synchronization latency.
Since latency hiding is effective in any type of network topology, we choose the ring topology. A ring minimizes the tail latency by simplifying the packet processing on the routers that would otherwise be substantial in a more complex network. In ESL architecture, each device is equipped with two quad small form-factor pluggable (QSFP) ports, supporting full-duplex communication for simultaneous transmission and reception.

%%%%%%%%%%%%%%%%%%%%%%%%%%%%%%%%%%%%%%%%%%%%%%%%%%
% Figures
%%%%%%%%%%%%%%%%%%%%%%%%%%%%%%%%%%%%%%%%%%%%%%%%%%
%%%%%%%%%%%%%%%%%%%%%%%%%%%%%%%%%%%%%%%%%%%%%%%%%%

\subsection{Reconfigurable Network} 
The necessity of a reconfigurable network in ESL arises from the requirement to efficiently support LLMs across various sizes and use cases. Even in an 8-device configuration, there are scenarios where operating with only 2 or 4 devices is more efficient. For instance, running two different models on a single server can be achieved by switching the model once inference is finished with one of the model on the 8-device configuration. However, the performance and efficiency loss occurs due to the switching overhead (e.g, model loading). Therefore, operating two models with two sets of 4 devices is more effective. To accommodate diverse model requirements, we implement a reconfigurable network in ESL to support 2, 4, and 8-device configuration as shown in Figure \ref{fig-esl}(b). In an 8-device configuration, a full ring is utilized, whereas in a 4-device configuration, it is split into two independent 4-lines. Similarly, a 2-device setup employs four 2-lines. The router determines the number and direction of hops based on the device ID to formulate a packet header that guarantees the most efficient communication path for synchronization. This design effectively maximizes the network resource without having to rewire or reload the model. Since each ring is guaranteed not to intersect with a different ring, each configuration can run independently to maximize the network resource within its own ring and thus the overall system. In all the configurations, the communication overhead is still the minimal tail latency, thereby achieving high scalability.

\section{HyperDex Framework}
We develop \textbf{HyperDex}, a software framework that consists of compilation layer and runtime layer, shown in Figure~\ref{fig-framework}(a), designed to deploy user applications and GenAI models to the LPU.
HyperDex's compilation layer and runtime layer are exposed as API to AI model architects and users, respectively.
These layers serve as a bridge, enabling both model architects and users to seamlessly integrate GenAI models and applications into LPU-equipped systems without requiring in-depth details about the underlying hardware.

\subsection{Compilation Layer}
The compilation layer within HyperDex framework hides the details of hardware and provides an abstracted interface of LPU to model architects, simplifying the deployment of GenAI models on LPU-equipped systems with requiring minimal effort.
We design the compilation layer to enable developers to import existing Huggingface models, and perform memory mapping, instruction generation, and compilation to generate binary program for the LPU hardware.
Moreover, model architects can customize, program, and compile their own models with API provided by HyperDex's compilation layer.
\textbf{HyperDex Model and Memory Mapper} analyzes the given model architecture and parameters, determining the most optimal memory allocation and alignment of each model parameter for maximum burst and streamlined processing within LPU.
HyperDex mapper considers system setups, such as number of devices and topology of the network, to partition the model parameters across multiple devices based on the intra-layer model parallelism, a type of model parallelism that divides the model parameters of parallelizable operations into multiple devices.
Since ESL innately complements model parallelism, its dataflow can stay transparent to HyperDex mapper.
Therefore, the mapper does not require further consideration to support ESL.
HyperDex model and memory mapper also takes into account memory (e.g., number of HBM channels and burst size) and compute (e.g., number of compute units) configurations to determine optimal memory mapping, tiling and padding size.
The mapper divides the multi-head attention weights with head-wise tiles and the feed-forward network weights with column-wise tiles, in which their dimensions are dependent to the hardware specifications.
The result is memory mapping of the tiled weights that perfectly matches the memory channel bitwidth and the order of operation to enable maximum utilization of memory bandwidth. 
The mapped parameters are fed to the instruction generator and loaded to LPU via HyperDex runtime layer.
%

%%%%%%%%%%%%%%%%%%%%%%%%%%%%%%%%%%%%%%%%%%%%%%%%%%
% Figures
%%%%%%%%%%%%%%%%%%%%%%%%%%%%%%%%%%%%%%%%%%%%%%%%%%
\begin{figure}[t] 
% \vspace{-0.01in}
\centering
\footnotesize
\includegraphics[width=1.0\linewidth]{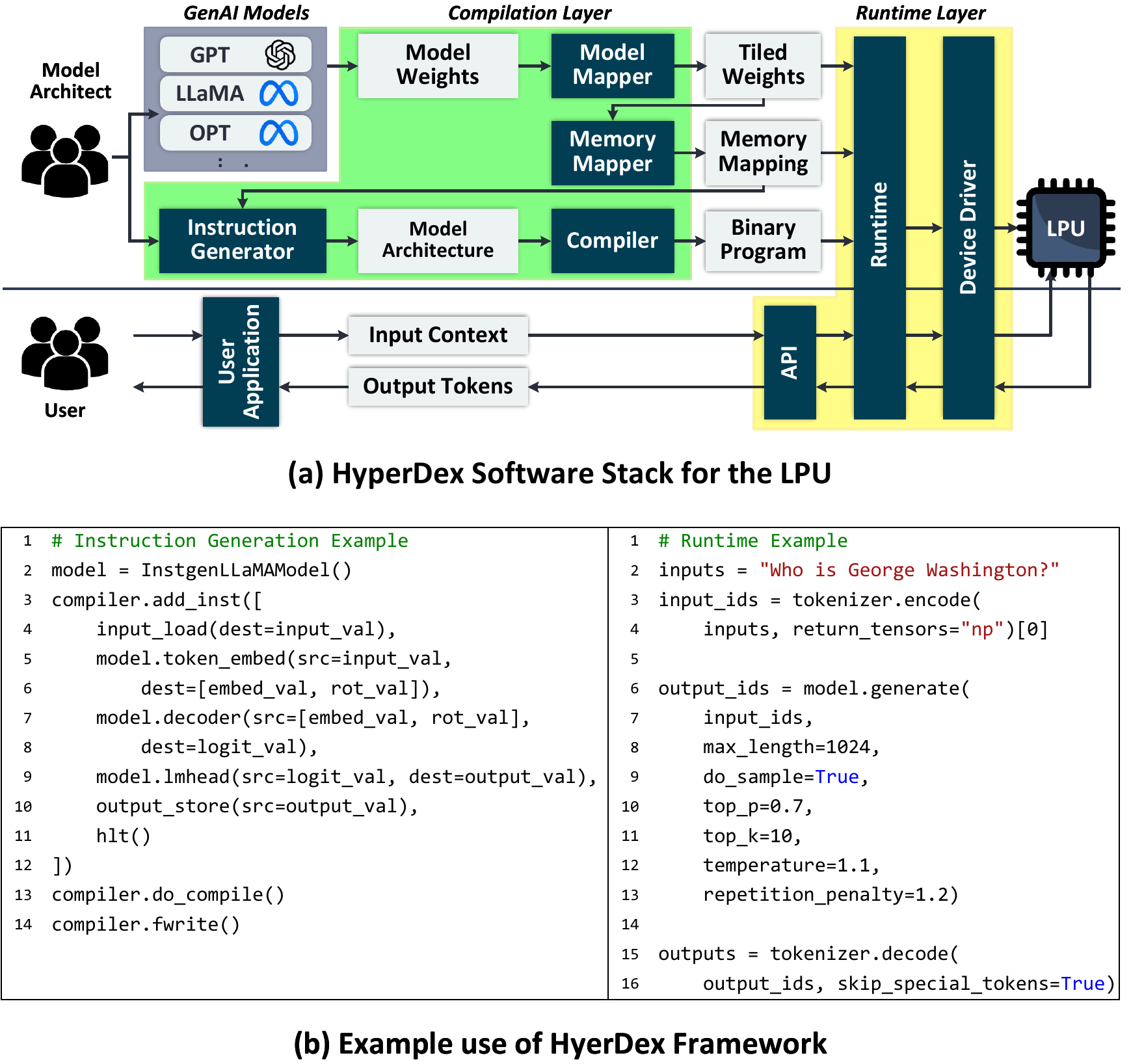} 
\vspace{-0.1in}
\caption{HyperDex software stack for the LPU.}
\label{fig-framework} 
\vspace{-0.2in}
\end{figure}
%%%%%%%%%%%%%%%%%%%%%%%%%%%%%%%%%%%%%%%%%%%%%%%%%%

%
\textbf{HyperDex Instruction Generator} creates a series of instructions for the LPU describing the GenAI model architecture based on the memory mapping information generated by HyperDex model and memory mapper.
HyperDex instruction generator provides a front-end that converts the Open Neural Network Exchange (ONNX) format into Python API calls of predefined instruction blocks for popular LLM models such as GPT, OPT, and Llama.
For instance, Figure~\ref{fig-framework}(b) demonstrates the model architecture code for the Llama model converted into Python with HyperDex instruction generator.
In this code snippet, the model architecture of Llama is defined through the following code blocks: 1) \texttt{input\_load} loads input user context via DMA and stores it in the designated variable, 2) \texttt{token\_embed}, \texttt{decoder}, and \texttt{lmhead} are predefined blocks encapsulating the core logic of the model, 3) \texttt{output\_store} returns the inferred output tokens via DMA, and 4) \texttt{hlt} marks the termination of the program.
Subsequently, this model architecture code is compiled with \texttt{do\_compile} function and is converted into a binary program through \texttt{fwrite} function.
For communicating between devices, \texttt{sync} block, composed of \texttt{transmit} and \texttt{receive} instructions, is utilized. 
As it is the role of the P2P interface to leverage ESL directly, the instruction generator operates independently of ESL.

For foundational models, the generation of the instruction blocks are automated, in which the parameters (e.g., source, destination, and their sizes) are parsed from the ONNX model.
For custom models, HyperDex provides API for model architects to program their own custom model architecture and corresponding operators using the instruction blocks.
Since LPU consists of a RISC-type processor and programmable engines, custom instruction blocks can be programmed and inserted to accommodate future operations and model architectures.
All of these blocks are translated into list of LPU instructions (i.e., machine codes).

\textbf{HyperDex Compiler} then compiles the instructions generated by HyperDex instruction generator into a binary program for the LPU hardware with several optimizations.
\textit{Register allocator} of the compiler tracks the lifetime of all variables and automatically allocates and releases the hardware registers at the compiler level.
It eliminates the necessity for the instruction generator and GenAI model architects to manually manage registers, offering them an abstracted view of the register file.
\textit{Instruction chaining} strategically divides the operations into a series of dependent instructions that can be executed back-to-back without any control overhead after initialization.
Our optimization for instruction chaining further separates instructions utilizing independent hardware modules into distinct groups (e.g., MEM, COMP, NET, CTRL) of instruction chains.
The compiler organizes a set of instructions performing specific functions from the distinct groups and interleave them so that the execution of each instruction can be overlapped.
It enables parallel execution of instruction chains across these groups, effectively minimizing the control overhead and latency.
\subsection{Runtime Layer}
Once the Information about the weights and architecture of the Llama model is loaded into the LPU, HyperDex runtime layer initializes the system. This initialization process occurs offline, and upon completion, other users can proceed to conduct model inference with the LPU-equipped system.
HyperDex's runtime layer provides a collection of API for user applications so that users can seamlessly integrate their GenAI applications with the LPU-based hardware.
The runtime layer provides API that align with the interfaces found in HuggingFace, including text generation, sampling, and streaming, allowing existing user applications to be integrated with LPU hardware with minimal code modification.
Figure~\ref{fig-framework}(b) demonstrates an example text generation application implemented using the API. Note that our design approach for the runtime API aligns with the interfaces of Huggingface. In this example code, \texttt{tokenizer} and \texttt{model} have the same interface as Huggingface's \texttt{AutoTokenizer} and \texttt{AutoModelForCausalLM}, respectively
HyperDex's runtime layer incorporates a device driver beneath the runtime API to perform low-level operations and hide complex hardware details from the developers.
This device driver extracts user-specified per-request and per-core arguments (e.g., core number, input and output token length, sampling parameters) via the runtime API and transferring this data to the control registers of LPU.
Whenever users send the requests through the runtime API, LPU receives these requests with user-specified arguments via the device driver, performs GenAI model inference, and returns the inference result.
Furthermore, HyperDex's runtime layer offers monitoring tools that provide hardware-level statistics such as power consumption, LPU utilization, and HBM usage, obtained from the device driver.
These tools are crucial in managing LPU-equipped systems at the datacenter level.

\section {Experimental Setup}

\subsection {ASIC Implementation}
\begin{figure}[!] 
\vspace{-0in}
\centering
\footnotesize
\includegraphics[width=0.9\linewidth]{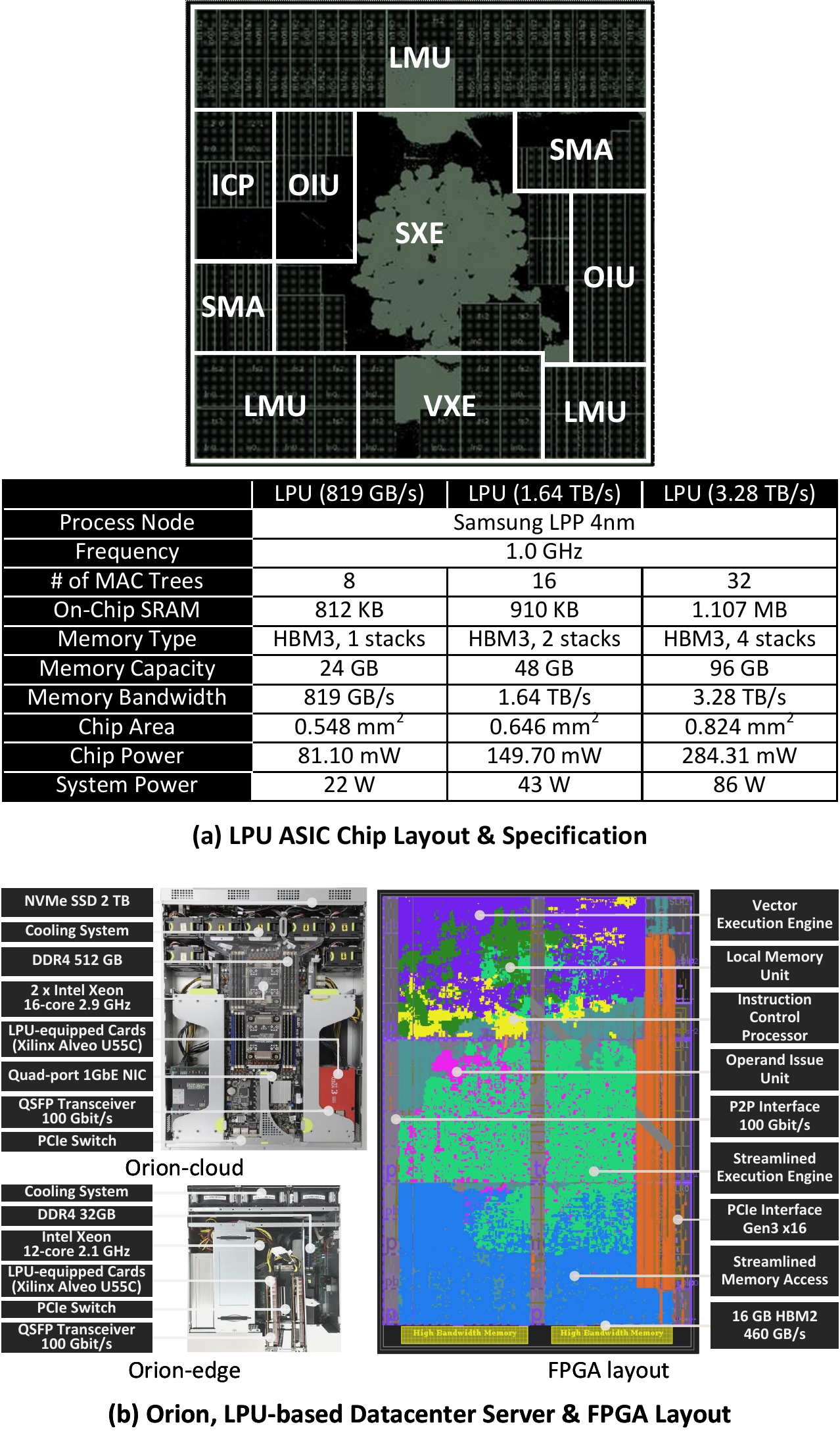} 
\vspace{-0.05in}
\caption{LPU implementation and specification.}
\label{fig-chip} 
\vspace{-0.20in}
\end{figure}

We develop RTL implementation of LPU using SystemVerilog and synthesize the proposed LPU architecture at a Samsung 4nm library and Synopsys Design Compiler. To demonstrate the scalability of our architecture, we synthesize the LPU with three different HBM configurations: 819 GB/s (24 GB), 1.64 TB/s (48 GB), and 3.28 TB/s (96 GB).  The vector dimension to fixed to 64 because the embedding dimension of most LLMs are multiples of 64, thus the number of MAC trees is set to 8, 16, and 32 to exactly match the memory bandwidths given the operating frequency of 1 GHz. An alternative is to scale down the vector dimension and proportionally scale up the number of MAC trees, but this configuration would halve the area of VXE at the cost of doubling its latency. We place and route using Synopsys IC Compiler II and measure the power and area through Synopsys PrimePower and Design Compiler tool, respectively. Figure \ref{fig-chip}(a) shows the overall LPU chip layout.

\textbf{ASIC simulator.} For performance evaluation, we implement in-house cycle-accurate simulator, written in C++, to measure the latency of LPU. It also simulates ESL to support any number of desired devices. We integrate ramulator with our simulator in order to simulate Samsung HBM3 Icebolt\textsuperscript{8}. We configure a single stack of HBM3 to have a speed of 819 GB/s and a capacity of 24 GB. In the evaluation, one, two, and four stacks of HBM3 are used for each HBM configuration. Also with the simulator, we observe that LPU occurs no accuracy loss on popular datasets (e.g., LAMBADA and Winograd Schema Challenge) with the open-source Megatron-LM GPT2, as the LPU supports the standard FP16 data precision.

%%%%%%%%%%%%%%%%%%%%%%%%%%%%%%%%%%%%%%%%%%%%%%%%%%
% Figures
%%%%%%%%%%%%%%%%%%%%%%%%%%%%%%%%%%%%%%%%%%%%%%%%%%
%%%%%%%%%%%%%%%%%%%%%%%%%%%%%%%%%%%%%%%%%%%%%%%%%%

\subsection {FPGA Implementation}
Before completing the fabrication of the LPU ASIC, we implement the LPU architecture on Xilinx Alveo U55C FPGA accelerator cards, where each device consists of HBM2 with 460 GB/s memory bandwidth and 16 GB capacity. Our FPGA implementation utilizes 46.4\% of LUT, 39.0\% of FF, 57.0\% of BRAM, 36.9\% of URAM, and 27.5\% of DSP, running at 220MHz kernel operating frequency. To match the memory bandwidth, LPU is configured with 16 MAC trees, which have total bandwidth of $16\times64\times2\,B\times220\,MHz\approx460\,GB/s$.

\textbf{Server-scale integration.} For initial commercialization, we productize HyperAccel Orion, a datacenter rack server based on the FPGA implementation, in two configurations: 1) Orion-cloud with eight LPU-equipped acceleration cards in a 2U server chassis with 128 GB and 3.3 TB/s HBM and 2) Orion-edge with two LPU-equipped acceleration cards in a edge server chassis with 32 GB and 960 GB/s HBM. For P2P communication, the devices are connected via ESL, a ring network connected by dual QSFP28 cables capable of 2$\times$100 Gbit/s. The software stack of Orion consists of the Xilinx Vitis 2022.2 platform and HyperDex framework with HuggingFace-like API for running multi-billion parameter LLMs. Figure \ref{fig-chip}(b) shows the implementation and server details.

\subsection {Methodology}
\textbf{LPU performance.} We measure the latency per output token of LPU with 3.28 TB/s HBM configuration. Since publicly available models on HuggingFace mostly consist of up to 70 billion parameters, we scale out to two LPUs to provide a total of 6.56 TB/s and 192 GB. For the model, we run widely-benchmarked OPT 1.3B, 6.7B, 30B, and 66B. Note that LPU supports other LLMs such as GPT, Llama, and their variants, but the latency is largely affected by the model size and not the type. For a fair performance comparison, we compare HyperAccel LPU with the state-of-the-art NVIDIA H100 GPU, which has comparable memory bandwidth of 3.35TB/s. We run LLM inference with the focus on generative tasks (e.g., article, code, and other text generation), in which we fix the number of input and output tokens to 32 and 2016, respectively. Note that the number of input tokens would barely affect the latency of GPU as it can process inputs in parallel.

\textbf{Server efficiency.} The efficiency of Orion is based on a real-system performance achieved while running an end-to-end text generation application. The number of tokens generated in one second per kilowatt of power consumption for each server is measured. We compare Orion-cloud and GPU server equipped with two NVIDIA H100s running OPT 1.3B to 66B and compare Orion-edge and GPU server equipped with two NVIDIA L4s running OPT 1.3B and 6.7B. The chosen models are variations of LLMs that fit into the given system. Note that memory space is labeled in decimal prefix (GB) but has physical capacity based on the binary prefix (GiB), so 66B model can fit into the 128 GB Orion-cloud system (e.g., 1.074 GB = 1 GiB). The compared systems have similar memory bandwidth specification and thermal design power (TDP).

\textbf{Scalability.} We compare the scalability of LPU and GPU in a multi-device setting. The GPU results are based on the open benchmark numbers of NVIDIA DGX A100 mentioned. DGX A100 is a server system that supports eight NVIDIA A100 GPUs on a 6U server chassis and third generation NVLink with GPU-to-GPU bandwidth of 600 GB/s. The scalability of LPU is validated with Orion-cloud with expandable synchronization link.

\section{Evaluation}

\textbf{Chip area and power analysis.} Figure \ref{fig-chip}(a) shows the area and power estimation breakdown of LPU in three different configurations. We specifically focus on the LPU with 32 MAC trees and 3.28 TB/s HBM, which has comparable memory bandwidth to the baseline NVIDIA H100 GPU with 3.35 TB/s HBM. The area and power consumption of the LPU chip is 0.824 mm\textsuperscript{2} and 284.31 mW, respectively. SXE dominates the area and power consumption of the LPU for housing majority of the compute logic, followed by SMA and LMU with mostly SRAMs for buffering and storing data. Including four HBM3 stacks, the total power consumption of the LPU system is 86 W. 
Compared to the H100 GPU, the LPU system requires only 15.2\% of the power consumption when running OPT 30B. The areas are not compared since GPU is a general-purpose processor and LPU is a domain-specific processor.

\subsection {LPU Performance}
\begin{figure}[t] 
\vspace{0in}
\centering
\footnotesize
\includegraphics[width=3.0in]{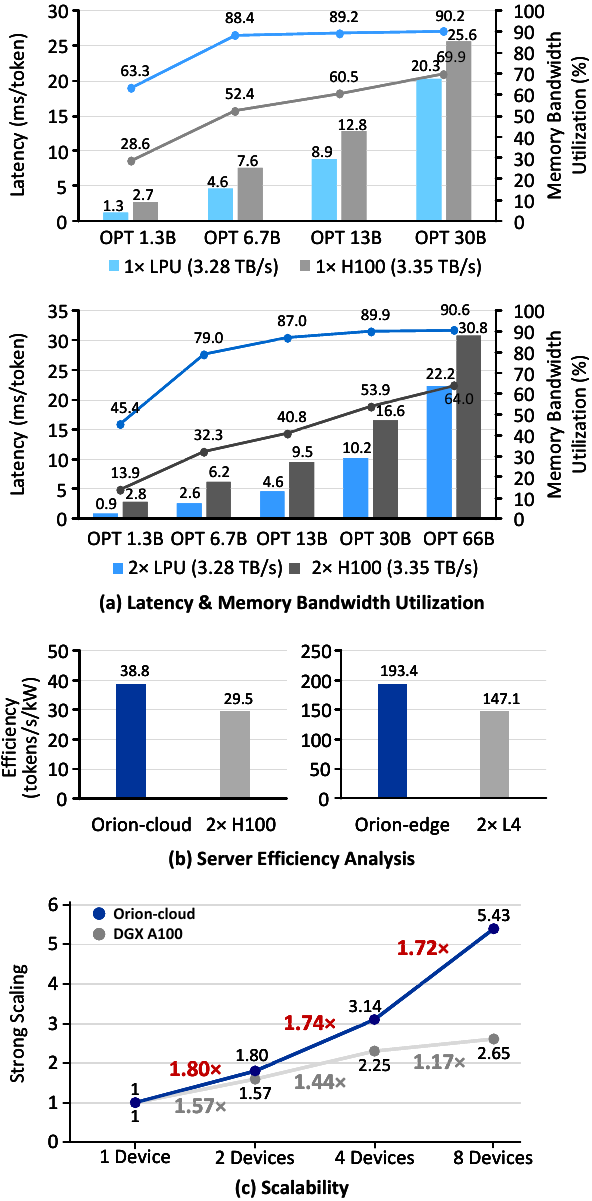} 
\vspace{-0.1in}
\caption{Performance of LPU compared to GPU.}
\vspace{-0.305in}
\label{fig-latencyasic} 
\end{figure}

\textbf{Latency.} Figure \ref{fig-latencyasic}(a) shows the simulated latency of LPU. Notably, LPU achieves 1.25 ms/token with 1.3B model and 4.62 ms/token with 6.7B model. For bigger OPT 66B model, two LPUs generate 1 token in just 22.2 ms. When compared with the equal number of H100, one LPU achieves 2.09$\times$ speedup on the 1.3B model, and two LPUs achieve 1.37$\times$ speedup on the 66B model. The streamlined architecture of LPU effectively uses the given memory bandwidth during the end-to-end inference compared to the GPU. On OPT 30B and 66B models, one LPU and two LPUs use up to 90.2\% and 90.6\% of the memory bandwidth, respectively, whereas one GPU and two GPUs use up to 70.8\% and 64.9\%, respectively. The advantage is more drastic for the smaller OPT 1.3B model, in which one LPU achieves 63.3\% memory bandwidth utilization, whereas one GPU achieves only 28.9\%.

%%%%%%%%%%%%%%%%%%%%%%%%%%%%%%%%%%%%%%%%%%%%%%%%%%
% Figures
%%%%%%%%%%%%%%%%%%%%%%%%%%%%%%%%%%%%%%%%%%%%%%%%%%
%%%%%%%%%%%%%%%%%%%%%%%%%%%%%%%%%%%%%%%%%%%%%%%%%%

\subsection {Server Efficiency}
Figure \ref{fig-latencyasic}(b) shows the efficiency analysis between HyperAccel Orion and GPU servers with comparable hardware specifications. For the cloud server, HyperAccel Orion-cloud with eight LPUs achieves 1.33$\times$ energy efficiency over the GPU server with 2$\times$ NVIDIA H100 when running OPT 66B model. Orion-cloud consumes 608 W, whereas H100 consumes 1100 W. For the edge server, Orion-edge with two LPUs achieves 1.32 $\times$ energy efficiency over the GPU server with 2$\times$ NVIDIA L4 when running OPT 6.7B model. Note that the efficiency advantage of device with LPU-based ASIC over the GPU would be significantly greater.

\subsection {Scalability}
LPU is specifically optimized to fully utilize its streamlined core for the small-batch computing required by LLM inference, whereas GPU suffers from severe underutilization for such workload. However, GPU still requires multiple devices for the best performance because 1) memory bandwidth is the bottleneck, and 2) additional memory capacity is required to support the larger models. The GPU undertuilization is accentuated with additional devices. Moreover, LPU devise a custom protocol to hide majority of the synchronization latency, while GPU undergoes significant overhead, as shown in the strong scaling of the two corresponding processors in Figure \ref{fig-latencyasic}(c). 
We analyze the scaling efficiency of up to 8 devices when running GPT3-20B. LPU achieves 5.43$\times$ speedup for the output token generation compared to a single device, which is significantly better than the 2.65$\times$ speedup of DGX A100. LPU achieves 1.75$\times$ speedup on average for doubling the number of devices due to the high scalability of the ESL technology, whereas GPU achieves only 1.38$\times$ speedup when the number of devices doubles due to the inability to hide the designated synchronization latency after the matrix multiplication. The scalability of LPU is verified with the Orion-cloud product.

\section{Related Work}
\label{related_work}

Recently, there has been a surge in the development of AI processors that accelerate LLM inference with competitive performance. 
Intel Habana Gaudi2 is a specialized accelerator designed for deep learning training and inference\textsuperscript{9}. It excels in handling small tensor operations, which facilitates simultaneous computation and network communication among diverse components. This capability also helps in reducing the bandwidth demands on its memory subsystem. 

Groq Language Processing Unit is based on the tensor streaming processor with optimizations to support LLM inference\textsuperscript{10}. It integrates 16 chip-to-chip interconnects and 230MB of SRAM, offering versatile options for embedded applications. However, the absence of external memory, such as HBM, requires Groq to integrate hundreds of cores to effectively accelerate practical LLMs with tens of billions of parameters (e.g., 512 chips for Llama2-70B), which incurs substantial communication overhead.

While these processors consist of architectures tailored for general AI computing to effectively handle LLM, HyperAccel LPU boasts an LLM-specific streamlined dataflow with high memory bandwidth and compute utilization to achieve unprecedented efficiency for LLM inference. Moreover, LPU features an efficient interconnect system that overlaps computation and communication between processors, which is beneficial for current LLMs that require systems ranging from node scale to rack scale.

\section{Conclusion}
% Rephrase all
We present a new class of processing unit, LPU, a latency-optimized and highly scalable architecture that accelerates large language model inference for generative AI. It introduces the streamlined processor and expandable synchronization link that maximize the bandwidth usage for agile processing and hide the synchronization overhead in peer-to-peer computing, respectively. HyperDex software framework assists the LPU to provide an optimized end-to-end solution. 

LPU-based ASIC achieves token generation latency of 1.25 ms/token for 1.3B model, and two LPUs achieve 20.9 ms/token for 66B model, while having a total area of 0.824 mm\textsuperscript{2} and power consumption of 284.31 mW. In addition, we implement LPU on cloud and edge FPGA servers to achieve 1.33$\times$ and 1.32$\times$ higher energy efficiency over NVIDIA H100 and L4 GPU server solutions, respectively.

Increasing the number of reused parameters would alleviate the memory bottleneck and proportionally increase the performance. Therefore, our future work includes developing an architecture that exploits the use of identical weights for different input contexts and batches, under the assumption that the operations are synchronized by layer. With additional sets of SXE and VXE, LPU can support two modes for parameter reuse. First, the multi-token mode that supports simultaneous execution of multiple input tokens would speedup the initial summarization stage. This mode can reduce the latency significantly for user requests with long input tokens. Second, batch mode that supports different user requests simultaneously would greatly improve the throughput, which is essential in high-traffic datacenters. The two modes would further increase the LPU performance while maintaining its outstanding efficiency and scalability.

In addition, we also consider a hybrid system composed of both GPU and LPU to cover a broader workload scope. Since GPU excels in other compute-intensive AI workloads (e.g., image processing) and LPU specializes in the sequential generation of textual content, combining the two systems should effectively handle multi-modal workloads, such as text-to-image generation. By overcoming the communication latency between GPU and LPU with dataflow optimizations, we expect that LPU and its area of application can expand to more diverse domains.

% \section{ACKNOWLEDGMENTS}
% The Acknowledgments is always plural even if there is a single acknowledgment. The author(s) would like to thank A, B, and C. This work was supported by XYZ under Grant \#\#\#.

% The ``Acknowledgments'' (spelled with just two e's, per American English) section appears immediately after the conclusion and before the reference list. Sponsor and financial support  are included in the acknowledgments section. For example: ``This work was supported in part by the U.S. Department of Commerce under Grant 123456.'' If support for a specific author is given, then use the following example for correct  wording. ``The work of First A. Author was supported by the U.S. Department of Commerce under Grant 123456''. Researchers that contributed information or assistance to the article should also be acknowledged in this section, and expressions should be simple and expressed as ``We thank$\ldots$,'' rather than indicating which of the authors is doing the thanking. Also, if corresponding authorship is noted in the paper, it should be placed in the bio of the corresponding author.

\def\refname{REFERENCES}

% \vspace*{-8pt}

\begin{IEEEbiography}{Seungjae Moon}{\,} is co-founder and hardware engineer at HyperAccel. His research interests include hardware architecture for machine learning model inference and dataflow optimization for memory-intensive applications. He received his M.S. in Electrical Engineering at KAIST. Contact him at sj.moon@hyperaccel.ai.
%\vadjust{\vfill\pagebreak}
\end{IEEEbiography}

\begin{IEEEbiography}{Jung-Hoon Kim}{\,} is a Ph. D. student at KAIST.
%is Co-founder and hardware engineer at HyperAccel. 
\end{IEEEbiography}

\begin{IEEEbiography}{Junsoo Kim}{\,} is co-founder and software engineer at HyperAccel. 
\end{IEEEbiography}

\begin{IEEEbiography}{Seongmin Hong}{\,} is co-founder and hardware engineer at HyperAccel. 
\end{IEEEbiography}

\begin{IEEEbiography}{Junseo Cha}{\,} is co-founder and hardware engineer at HyperAccel. 
\end{IEEEbiography}

\begin{IEEEbiography}{Minsu Kim}{\,} is a Ph.D. student at KAIST. 
\end{IEEEbiography}

\begin{IEEEbiography}{Sukbin Lim}{\,} is a hardware engineer at HyperAccel. 
\end{IEEEbiography}

\begin{IEEEbiography}{Gyubin Choi}{\,} is a software engineer at HyperAccel. 
\end{IEEEbiography}

\begin{IEEEbiography}{Dongjin Seo}{\,} is a hardware engineer at HyperAccel. 
\end{IEEEbiography}

\begin{IEEEbiography}{Jongho Kim}{\,} is a hardware engineer at HyperAccel. 
\end{IEEEbiography}

\begin{IEEEbiography}{Hunjong Lee}{\,} is a software engineer at HyperAccel. 
\end{IEEEbiography}

\begin{IEEEbiography}{Hyunjun Park}{\,} is a software engineer at HyperAccel. 
\end{IEEEbiography}

\begin{IEEEbiography}{Ryeowook Ko}{\,} is a hardware intern at HyperAccel. 
\end{IEEEbiography}

\begin{IEEEbiography}{Soongyu Choi}{\,} is a hardware intern at HyperAccel. 
\end{IEEEbiography}

\begin{IEEEbiography}{Jongse Park}{\,} is an Associate Professor in the School of Computing at KAIST. Contact him at jspark@casys.kaist.ac.kr. 
\end{IEEEbiography}

\begin{IEEEbiography}{Jinwon Lee}{\,} is CTO of HyperAccel. Contact him at jw.lee@hyperaccel.ai. 
\end{IEEEbiography}

\begin{IEEEbiography}{Joo-Young Kim}{\,} is CEO of HyperAccel. Contact him at jy.kim@hyperaccel.ai.
\end{IEEEbiography}

% \begin{IEEEbiography}{Third C. Author III} {\,} is a program officer at the  DEF Corporation, Tokyo, Japan. Contact him at tcauthor@def.com.
% \end{IEEEbiography}


\begin{thebibliography}{1}

\bibitem{AA1}
A.Vaswani et al., ``Attention is all you need,'' {\it 31st Conference on Neural Information Processing Systems (NIPS)}., pp. 6000-–6010, 2017, doi: 10.48550/arXiv.1706.03762

\bibitem{BB1}
Chowdhery et al., ``Palm: Scaling language modeling with pathways,'' {\it Journal of Machine Learning Research (JMLR)}., vol. 24, no. 1, pp. 11324--11436, 2023, doi: 10.48550/arXiv.2204.02311

\bibitem{CC1}
H. Wang, Z. Zhang and S. Han, ``SpAtten: Efficient Sparse Attention Architecture with Cascade Token and Head Pruning,'' {\it 2021 IEEE International Symposium on High-Performance Computer Architecture (HPCA)}., pp 97--110, 2021. doi: 10.1109/HPCA51647.2021.00018

\bibitem{DD1}
Wolf et al., ``Transformers: State-of-the-art natural language processing,'' {\it Proceedings of the 2020 Conference on Empirical Methods in Natural Language Processing: System Demonstrations (EMNLP)}., pp. 38--45, 2020, doi: 10.48550/arXiv.1910.03771

\bibitem{EE1}
Hong et al., ``DFX: A Low-latency Multi-FPGA Appliance for Accelerating Transformer-based Text Generation,'' {\it 2022 55th IEEE/ACM International Symposium on Microarchitecture (MICRO)}., pp. 616--630, 2022, doi: 10.1109/MICRO56248.2022.00051.

\bibitem{FF1}
Song et al., ``On-the-fly operation batching in dynamic computation graphs,'' {\it 31st Conference on Neural Information Processing Systems (NIPS)}., pp. 3974–-3984, 2017, doi: 10.48550/arXiv.1706.03762.

\bibitem{GG1}
N. Kitaev, L. Kaiser, A. Levskaya., ``Reformer: The Efficient Transformer,'' {\it International Conference on Learning Representations (ICLR)}., doi: 10.48550/arXiv.2001.04451

% @article{kitaev2020reformer,
%   title={Reformer: The efficient transformer},
%   author={Kitaev, Nikita and Kaiser, {\L}ukasz and Levskaya, Anselm},
%   journal={arXiv preprint arXiv:2001.04451},
%   year={2020}
% }

\bibitem{HH1}
Y. Kim, W. Yang, and O. Mutlu, ``Ramulator: A fast and extensible DRAM simulator,'' {\it IEEE Computer Architecture Letters}., vol. 15, no. 1, pp. 45-49, 2015 doi: 10.1109/LCA.2015.2414456.

\bibitem{II1}
E. Medina and E. Dagan, ``Habana labs purpose-built ai inference and training processor architectures: Scaling ai training systems using standard ethernet with gaudi processor,'' {\it IEEE Micro}., vol. 40, no. 2, pp. 17--24, 2015 doi: 10.1109/MM.2020.2975185.

\bibitem{JJ1}
D. Abts et al., ``A software-defined tensor streaming multiprocessor for large-scale machine learning,'' {\it Proceedings of the 49th Annual International Symposium on Computer Architecture (ISCA)}., pp. 567--580, 2022 doi: 10.1145/3470496.3527405.

\end{thebibliography}
\end{document}